\documentclass[12pt,transaction, draftclsnofoot,onecolumn]{IEEEtran}
\usepackage{amsfonts}
\usepackage{amssymb}
\usepackage{graphicx,epstopdf}
\usepackage{subfigure}
\usepackage{enumerate}
\usepackage{amsmath}
\usepackage{color}
\usepackage{amsthm}
\usepackage{amsmath}
\usepackage{algorithm}
\usepackage{algpseudocode}
\usepackage{cite}

\newcommand{\bp}{\begin{proof} \small }
\newcommand{\ep}{\end{proof} \normalsize}
\newcommand{\epx}{\end{proof} \small}
\newcommand{\bpa}{\begin{proofappx} \footnotesize }
\newcommand{\epa}{\end{proofappx} \small }
\newtheorem{theorem}{Theorem}

\newtheorem*{theorem*}{Theorem}
\newtheorem*{proposition*}{Proposition}
\newtheorem*{corollary*}{Corollary}
\newtheorem*{lemma*}{Lemma}
\newtheorem*{assumption*}{Assumption}
\newtheorem*{definition*}{Definition}
\newtheorem*{claim*}{Claim}

\newcommand{\be}{\begin{equation}}
\newcommand{\ee}{\end{equation}}
\newcommand{\bs}{\begin{subequations}}
\newcommand{\es}{\end{subequations}}
\newcommand{\bq}{\begin{eqnarray}}
\newcommand{\eq}{\end{eqnarray}}
\newcommand{\bqn}{\begin{eqnarray*}}
\newcommand{\eqn}{\end{eqnarray*}}

\newcommand{\ba}{\left[ \begin{array}}
\newcommand{\ea}{\\ \end{array} \right]}
\newcommand{\ben}{\begin{enumerate}}
\newcommand{\een}{\end{enumerate}}

\def\real{{\mathchoice%
{\hbox{\rm\setbox1=\hbox{I}\copy1\kern-.45\wd1 R}}
{\hbox{\rm\setbox1=\hbox{I}\copy1\kern-.45\wd1 R}}
{\hbox{\scriptsize\rm\setbox1=\hbox{I}\copy1\kern-.45\wd1 R}}
{\hbox{\scriptsize\rm\setbox1=\hbox{I}\copy1\kern-.45\wd1 R}}}}

\def\Zint{{\mathchoice{\setbox1=\hbox{\sf Z}\copy1\kern-.75\wd1\box1}
{\setbox1=\hbox{\sf Z}\copy1\kern-.75\wd1\box1}
{\setbox1=\hbox{\scriptsize\sf Z}\copy1\kern-.75\wd1\box1}
{\setbox1=\hbox{\scriptsize\sf Z}\copy1\kern-.75\wd1\box1}}}
\newcommand{\complex}{ \hbox{\rm C\kern-0.45em\rule[.07em]{.02em}{.58em}%
\kern 0.43em}}

\IEEEoverridecommandlockouts

\begin{document}
%
\title{Online Learning for Offloading and Autoscaling in Renewable-Powered Mobile Edge Computing}

\author{
\IEEEauthorblockN{Jie Xu,}
\IEEEauthorblockA{University of Miami\\}
\and
\IEEEauthorblockN{Shaolei Ren,}
\IEEEauthorblockA{University of California, Riverside}
}


%


\maketitle

\begin{abstract}
Mobile edge computing (a.k.a. fog computing) has recently emerged
to enable \emph{in-situ} processing of delay-sensitive applications
at the edge of mobile networks. Providing grid power supply in support
of mobile edge computing, however, is costly and even infeasible (in certain
rugged or under-developed areas),
thus mandating on-site renewable energy as a major
or even sole power supply in increasingly many scenarios.
Nonetheless, the high intermittency and unpredictability of renewable energy make it very challenging to deliver a high quality of service to users
in renewable-powered mobile edge computing systems.
In this paper, we address the challenge of incorporating renewables
into mobile edge computing and
propose an efficient reinforcement learning-based resource management algorithm,
which learns on-the-fly the optimal policy of
dynamic workload offloading (to centralized cloud) and
edge server provisioning to minimize the long-term system cost
(including both service delay and operational cost).
Our online learning algorithm uses a decomposition of the (offline) value iteration and (online) reinforcement learning, thus achieving a significant improvement of
learning rate and run-time performance when
compared
to standard reinforcement learning algorithms such as Q-learning.


\end{abstract}

%
\IEEEpeerreviewmaketitle

\section{Introduction}

In the era of mobile computing and Internet of Things,
a tremendous amount of data is generated from
massively distributed sources, requiring timely processing to extract its maximum value. Further,
many emerging applications, such as mobile gaming and augmented reality,
are delay sensitive and have resulted in an increasingly high computing demand
that frequently exceeds what mobile devices can deliver.
Although cloud computing enables convenient access to a centralized pool of configurable computing resources, moving all the distributed data and computing-intensive applications
to clouds (which are often physically located in remote mega-scale data centers)
is simply out of the question, as it would not only pose
 an extremely heavy burden on today's already-congested
 backbone networks but also result in (sometimes intolerable) large transmission
 latencies that degrade the quality of service.

As a remedy to the above limitations, mobile edge computing (a.k.a., fog computing)
has recently emerged to enable \emph{in-situ} processing of
(some) workloads locally at the network edge without moving them to the cloud \cite{beck2014mobile,vaquero2014finding}. In mobile
edge computing,
network edge devices, such as base stations, access points and routers,
are empowered with computing and storage capabilities to serve users' requests
as a substitute of clouds, while significantly reducing the transmission latency
as they are placed in the proximity of end users.
In this paper, we consider (marco) base station as the default edge device
and refer to the combination of an edge device and the associated edge servers as an {\it edge system}.

\begin{figure}
  \centering
  \includegraphics[scale = 0.6]{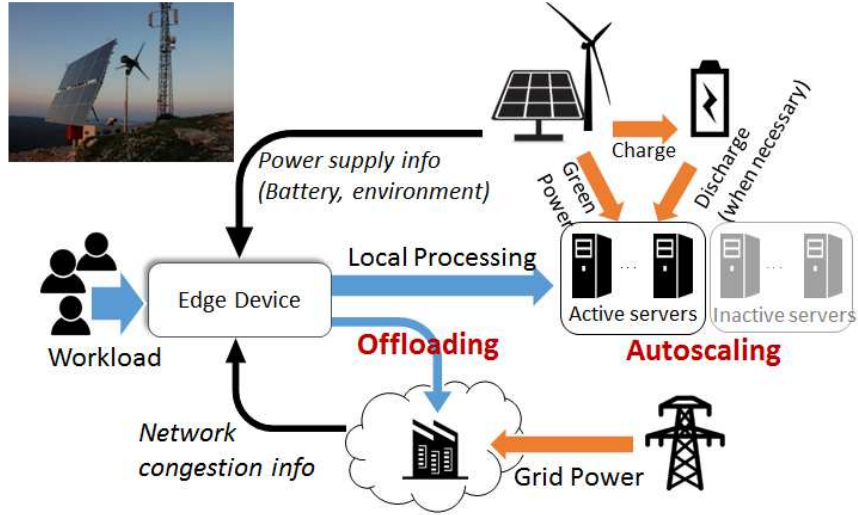}\\
  \caption{Architecture of a renewable-powered edge computing system. The photo shows a solar- and wind-powered system deployed by Alcatel Lucent in Turkey. (Source: http://www.cellular-news.com/tags/solar/wind-power/)}\label{system}
\end{figure}

In increasingly many scenarios, edge systems are primarily powered by renewable green energy (e.g. solar and wind), rather than the conventional electric grid, due to various reasons such as location, reliability, carbon footprint and cost. For instance, in many developing countries, the majority of base stations have to be powered by continuously operating diesel generators because the electric grid is too unreliable \cite{TaoHan_UNCC_LoadBalancingRAN_HybriedEnergy_ToN}. Even if power line extension is technically feasible, grid-tied edge systems can violate environmental quality regulations in rural areas that are ecologically sensitive. Thus, in view of the significant carbon footprint of grid power as well as soaring electricity prices, renewable energy is embraced as a major energy source. Despite the clear advantages,
a distinct feature of renewable energy is that it can vary drastically over time and is highly unpredictable. Although batteries are often installed as an energy buffer, the computing capacity of an edge system is still significantly limited at any moment in time. As a result, although edge processing reduces the transmission latency, a considerable processing time may occur when little power supply is available. This gives rise to an important trade-off between transmission delay and processing delay, which is jointly determined by the edge system's offloading policy (i.e. how much workload is offloaded to the cloud) and autoscaling policy (i.e. how many servers are dynamically provisioned or activated). The problem is further complicated due to the temporal correlation --- provisioning more servers and processing more workloads at the edge system in the current time means that fewer servers can be provisioned and fewer workloads can be processed locally in the future due to the limited and time-varying renewable energy supply. Figure \ref{system} illustrates the considered system architecture.

In this paper, we address the challenge of incorporating renewables
into mobile edge computing and
propose an efficient reinforcement learning-based resource management algorithm,
which learns on-the-fly the optimal policy of
dynamic workload offloading (to centralized cloud) and
edge server provisioning to minimize the long-term system cost
(including both service delay and operational cost).
The problem is formulated as a Markov decision process (MDP) by taking into account various unique aspects of the considered mobile edge system. A novel post-decision state (PDS) based algorithm that learns the optimal joint offloading and autoscaling policy on-the-fly is developed. Compared with conventional online reinforcement learning algorithms, e.g. Q-learning, the proposed PDS based learning algorithm significantly improves the convergence speed and run-time performance by exploiting the special structure of the considered problem. Based on extensive simulations and numerical results, we show that our algorithm can significantly improve the performance of the green mobile edge computing system.

\section{Related Work}
Mobile edge computing has received an increasing amount
of attention in recent years. In particular,
a central theme of many prior studies is offloading policy on
the \emph{user} side, i.e.,
what/when/how to offload a user's workload
from its device to the edge system or cloud
(see \cite{Offloading_HuangWangNiyato_TWC_2012,Cloudlet_CMU_Satyanarayanan:2009:CVC:1638591.1638731}
and references therein). Our work on edge-side offloading
and autoscaling
is complementary to these studies on user-side offloading.

Our study is also relevant to the rich literature on power management
in wireless networks \cite{chia2014data,USC_DynamicBaseStation_Switching_On_Off_TWireless_2013,Krishnamachari_USC_Energy_efficient_wireless_basestation_CommunicationsMagazine_2011}
and data centers \cite{LinWiermanAndrewThereska}, especially in the context of renewable-powered systems \cite{TaoHan_UNCC_LoadBalancingRAN_HybriedEnergy_ToN,ChaoLi_iSwitch_ISCA_2012_Li:2012:ICO:2337159.2337218,Goiri:2011:GSE:2063384.2063411}.
Nonetheless, our study differs from these works
as we jointly optimize the offloading and autoscaling
decisions at the edge system, whereas prior research
on edge device (base station) or data center power
management typically only considers one of the two decisions.
For example, autoscaling (a.k.a., right-sizing) in data centers \cite{LinWiermanAndrewThereska}
dynamically controls the number of active servers, but the control
knob of offloading to the cloud is not available in the context of data centers.
While some studies on base station power management
considers traffic offloading to small cells and/or other base stations \cite{TaoHan_UNCC_LoadBalancingRAN_HybriedEnergy_ToN}, we study an
orthogonal type of offloading --- from edge servers to the cloud --- which
requires different models and techniques (see Section~III-D).
Further, in contrast with these studies
\cite{chia2014data,TaoHan_UNCC_LoadBalancingRAN_HybriedEnergy_ToN,ChaoLi_iSwitch_ISCA_2012_Li:2012:ICO:2337159.2337218,Goiri:2011:GSE:2063384.2063411,LinWiermanAndrewThereska,USC_DynamicBaseStation_Switching_On_Off_TWireless_2013,Krishnamachari_USC_Energy_efficient_wireless_basestation_CommunicationsMagazine_2011},
we propose a novel solution technique based on reinforcement learning
to incorporate intermittent and unpredictable renewables
into mobile edge systems. Finally, we note that the most relevant study to our work is \cite{deng2015towards},
which also studies workload allocation/offloading in a cloud-fog computing system. However, unlike
our renewable-powered edge system,
this paper considers a grid-powered system and focuses on a one-shot static optimization
without addressing the temporal correlation among the offloading decisions across time (due
to intermittent renewables and limited battery).

\section{System Model}

As a major deployment method of mobile edge computing \cite{MEC_ESTI_WhitePaper_2014},
we consider an edge system consisting of a base station
and a set of edge servers, which are physically co-located and share
the same power supply in
the cell site.

\subsection{Workload model}
We consider a discrete-time model by dividing the operating period into time slots of equal length indexed by $t=0,1,...$, each of which has a duration that matches the timescale at which the edge device can adjust its computing capacity (i.e. number of active servers). We use $x \in \mathcal{L}$ to represent a location coordinate in the service area $\mathcal{L}$. Let $\lambda(x,t)$ represent the workload arrival rate in location $x$, and $\theta(x, t)$ be the wireless transmission rate between the base station and location $x$. Thus $\lambda(t) = \sum_{x\in\mathcal{L}} \lambda(x,t) \in [0, \lambda_{max}]$ is the total workload arrival rate at the edge system, where $\lambda_{max}$ is the maximum possible arrival rate. The system decides the amount of workload $\mu(t) \leq \lambda(t)$ that will be processed locally. The remaining workload $\nu(t) \triangleq \lambda(t) - \mu(t)$ will be offloaded to the cloud for processing. The edge system also decides at the beginning of the time slot the number of active servers, denoted by $m(t) \in [0, M] \triangleq \mathcal{M}$. These servers are used to serve the local workload $\mu(t)$. Since changing the number of servers during job execution are difficult and in many cases impossible, we only allow determining the number of servers at the beginning of each time slot but not during the slot.

\subsection{Power model}
We interchangeably use power and energy, since energy
consumption during each time slot is
the product of (average) power and the duration of each time slot that
is held constant in our model.
The total power demand of the edge system in a time slot consists of two parts: first, basic operation and transmission power demand by edge devices (base station in our study); and second, computing power demand by
edge servers. The first part is independent of the offloading or the autoscaling policy, which is modeled as $d_{op}(\lambda(t)) = d_{sta} + d_{dyn}(\lambda(t))$
where $d_{sta}$ is the static power consumption and $d_{dyn}(\lambda(t))$ is the dynamic power consumption depending on the amount of total workload. The computing power demand depends on the number of active servers as well as the locally processed workload. We use a generic function $d_{com}(m(t), \mu(t))$, which is increasing in $m(t)$ and $\mu(t)$, to denote the computing power demand. The total power demand is therefore
\begin{align}
d(\lambda(t), m(t), \mu(t)) = d_{op}(\lambda(t)) + d_{com}(m(t), \mu(t))
\end{align}

To model the uncertainty of the green power supply, we assume that the green power budget, denoted by $g(t)$, is realized after the offloading and autoscaling decisions are made. Therefore, the decisions cannot utilize the exact information of $g(t)$. However, we assume that there is an environment state $e(t)$ which the system can observe and it encodes valuable information of how much green energy budget is anticipated in the current time slot. For instance, daytime in a good weather usually implies high solar power budget. Specifically, we model $g(t)$ as an i.i.d. random variable given $e(t)$, which obeys a conditional probability distribution $P_g(g(t)|e(t))$. Note that the environment state $e(t)$ itself may not be i.i.d.

\subsection{Battery model}
Batteries are used to balance the power supply and demand. In a solar+wind system, photovoltaic modules and wind turbines can combine their output to power the edge system and charge the batteries. When their combined efforts are insufficient, batteries take over to ensure steady operation of the edge system. We denote the battery state at the beginning of time slot $t$ by $b(t) \in [0, B] \triangleq \mathcal{B}$ (in units of power) where $B$ is the battery capacity. For system protection reasons, the battery unit has to be disconnected from the load once its terminal voltage is below a certain threshold for charging. We map $b(t) = 0$ to this threshold voltage to ensure basic operation of the system. Since green power budget is unpredictable and hence unknown at the beginning of time slot $t$, the edge system uses a conservative policy which satisfies $d_{com}(m(t), \mu(t)) \leq \max\{b(t) - d_{op}(\lambda(t)),0\}$. It instructs the edge system to offload all workload to the cloud if the existing battery level cannot even support the basic operation and transmission in the current slot. When $d_{op}(\lambda(t)) \geq b(t)$, the backup power supply (e.g. diesel generator) will be used to maintain basic operation for the slot. The cost due to activating the backup power supply is $
c_{bak}(t) = \phi\cdot d_{op}(\lambda(t))$ where $\phi > 0$ is a large constant representing the large cost due to using the backup power supply. The next time slot battery state then evolves to $b(t+1) = b(t) + g(t)$. When $d_{op}(\lambda(t)) \leq b(t)$, the edge system may process part of the workload $\mu(t) \leq \lambda(t)$ at the local servers. Depending on the realized green power $g(t)$ and the computing power demand $d_{com}(\lambda(t), m(t), \mu(t))$, the battery is recharged or discharged accordingly:
\begin{itemize}
  \item  If $g(t) \geq d(\lambda(t), m(t), \mu(t))$, then the surplus $g(t) - d(\lambda(t), m(t), \mu(t))$ is stored in the battery until reaching its capacity $B$.
  \item If $g(t) < d(\lambda(t), m(t), \mu(t))$, then the battery is discharged to cover the deficit $d(\lambda(t), m(t), \mu(t)) - g(t)$.
\end{itemize}

For simplicity, we will assume that there is no power loss either in recharging or discharging the batteries, noting that this can be easily generalized. We also assume that the batteries are not leaky. We model the battery depreciation cost in a time slot, denoted by $c_{battery}(t)$, using the amount of discharged power in this time slot since the lifetime discharging is often limited. Specifically,
\begin{align*}
c_{battery}(t) = \omega\cdot\max\{d(\lambda(t), m(t), \mu(t)) - g(t),0\}
\end{align*}
where $\omega > 0$ is the normalized unit depreciation cost.

\subsection{Delay cost model}
The average utilization of the base station is $\rho(t) = \sum_x \lambda(x,t)/\theta(x,t)$,
which results in a total wireless access and transmission delay of $c_{wi}(t) = \sum_{x} \lambda(x,t)/[\theta(x,t)(1-\rho(t)]$ by following the literature and
modeling the base
station as a queueing system \cite{USC_DynamicBaseStation_Switching_On_Off_TWireless_2013}. Next we model the workload processing delay incurred at the edge servers.

For the local processed workload, the delay cost $c_{lo}(t)$ is mainly processing delay due to the limited computing capacity at the local edge servers. The transmission delay from the edge device to the local servers is negligible due to physical co-location. To quantify the delay performance of services, such as average delay and tail delay (e.g. 95th-percentile latency), without restricting our model to any particular performance metric, we use the general notion of $c_{lo}(m(t), \mu(t))$ to represent the delay performance of interest during time slot $t$. As a concrete example, we can model the service process at a server instance as an M/G/1 queue and use the average response time (multiplied by the arrival rate) to represent the delay cost, which can be expressed as $c_{lo}(m(t), \mu(t)) = \frac{\mu(t)}{m(t)-\mu(t)}$.

For the offloaded workload, the delay cost $c_{off}(t)$ is mainly transmission delay due to network round trip time (RTT), which varies depending on the network congestion state. For
modeling simplicity, the service delay at the cloud side is also absorbed
into the network congestion state.
Thus, we model the network congestion state, denoted by $h(t)$, as an exogenous parameter and express it in terms of the RTT (plus cloud service delay) for simplicity. The delay cost is thus $c_{off}(h(t), \lambda(t), \mu(t)) = (\lambda(t) - \mu(t))\max\{h(t)-d_0,0\}$. The total delay cost is therefore
\begin{align}
&c_{delay}(h(t), \lambda(t), m(t), \mu(t)) \nonumber\\
= &c_{lo}(m(t), \mu(t)) + c_{off}(h(t), \lambda(t), \mu(t)) + c_{wi}(\lambda(t))
\end{align}

\section{Problem Formulation}
In this section, we formulate the dynamic offloading and autoscaling problem as an online learning problem, in order to minimize the system cost. The system system is described by a tuple $s(t) \triangleq (\lambda(t), e(t), h(t), b(t))$, which is observable at the beginning of the time slot. Among the four state elements, $\lambda(t)$, $e(t)$, $h(t)$ are exogenous states which are independent of the offloading and autoscaling actions. To make the stochastic control problem tractable, they are assumed to have finite value spaces and evolve as finite-state Markov chains. Specifically, let $P_\lambda(\lambda(t+1)|\lambda(t))$, $P_e(e(t+1)|e(t))$ and $P_h(h(t+1)|h(t))$ denote the transition matrices for $\lambda(t)$, $e(t)$ and $h(t)$, respectively. Similar assumptions have been made in existing literature, e.g. \cite{GuenterJainWilliams}. Importantly, all these probability distributions are unknown {\it a priori} to the edge system.

The stochastic control problem now can be cast into an MDP, which consists of four elements: the state space $\mathcal{S}$, the action space $\mathcal{A}$, the state transition probabilities $P_s(s(t+1)|s(t), a(t)), \forall s,s'\in\mathcal{S}, a\in \mathcal{A}$, and the cost function $c(s,a), \forall s, a$. We have already defined the state space. Next we introduce the other elements as follows.

\textbf{Actions}. Although the actual actions taken by the edge system are $\nu(t)$ (offloading) and $m(t)$ (autoscaling) in each time slot $t$, we will consider an intermediate action in the MDP formulation, which is the computing power demand in each time slot $t$, denoted by $a(t) \in \mathcal{A}$ where $\mathcal{A}$ is a finite value space. We will see in a moment how to determine the optimal offloading and autoscaling actions based on this. As mentioned before, to maintain basic operation in the worst case, we require that $a(t) \leq \max\{b(t) - d_{op}(\lambda(t)), 0\}$.

\textbf{State transition probability}. Given the current state $s(t)$, the computing power demand $a(t)$ and the realized green power budget $g(t)$, the buffer state in the next time slot is
\begin{align}\label{buffer}
&b(t+1) = [b(t) + g(t)]_0^B, \textrm{if}~d_{op}(\lambda(t)) > b(t)\\
&b(t+1) = [b(t) - P_1(\lambda(t)) - a(t) + g(t)]_0^B, \textrm{otherwise} \nonumber
\end{align}
where $[\cdot]_0^B$ denotes $\max\{\min\{\cdot, B\}, 0\}$. The system then evolves into the next time slot $t+1$ with the new state $s(t+1)$. The transition probability from $s(t)$ to $s(t+1)$, given  $a(t)$, can be expressed as follows
\begin{align}
&P(s(t+1)|s(t),a(t)) \nonumber\\
=&P_\lambda(\lambda(t+1)|\lambda(t))P_e(e(t+1)|e(t))P_h(h(t+1)|h(t)) \nonumber\\
\times &\sum\limits_{g(t)}P_g(g(t)|e(t))\textbf{1}\{\zeta(t)\}
\end{align}
where $\textbf{1}\{\cdot\}$ is the indicator function and $\zeta(t)$ denotes the event defined by \eqref{buffer}. Notice that the state transition only depends on  $a(t)$ but not  the offloading or the autoscaling action. This is why we can focus on the computing power demand action $a(t)$ for the foresighted optimization problem.

\textbf{Cost function}. The total system cost is the sum of the delay cost, the battery depreciation cost and the backup power supply cost. If $d_{op}(\lambda(t)) > b(t)$, then the cost is simply
\begin{align}
\tilde{c}(s(t), a(t)) = c_{delay}(h(t), \lambda(t), 0, 0) + c_{bak}(\lambda(t))
\end{align}
since we must have $m(t) = 0$ and $\mu(t) = 0$. Otherwise, the realized cost given the realized green power budget $g(t)$ is
\begin{align*}
\tilde{c}(t) =  c_{delay}(h(t), \lambda(t), m(t), \mu(t))
+  \omega\cdot [a(t) - g(t)]_0^\infty
\end{align*}
Since the state transition does not depend on $\mu(t)$ or $m(t)$, they can be  optimized given $s(t)$ and $a(t)$ by solving the following myopic optimization problem
\begin{align}\label{joint}
\max_{\lambda_1\leq \lambda, m}~~ c_{delay}(h, \lambda, m, \mu)~~\textrm{s.t.}~~ P(m, \mu) = a
\end{align}
Let $m^*(s, a)$ and $\mu^*(s, a)$ denote the optimal solution and $c^*_{delay}(s, a)$ the optimal value given $s$ and $a$. Therefore, the minimum cost in time slot $t$ given $s$ and $a$ is
\begin{align*}
\tilde{c}(s(t), a(t), g(t)) = c^*_{delay}(s(t), a(t))+ \omega\cdot [a(t) - g(t)]_0^\infty
\end{align*}
The expected cost is thus
\begin{align*}
c(s(t), a(t)) = c^*_{delay}(s(t), a(t)) +  E_{g(t)|e(t)}\omega\cdot [a(t) - g(t)]_0^\infty
\end{align*}

\textbf{Policy}. The edge system's computing power demand policy (which implies the joint offloading and autoscaling policy) in the MDP is a mapping $\pi: \Lambda\times \mathcal{E}\times \mathcal{H} \times \mathcal{B} \to \mathcal{A}$. We focus on optimizing the policy to minimize the edge system's expected long-term cost, which is defined as the expectation of the discounted sum of the edge device's one-slot cost: $C^\pi(s(0)) = \mathbb{E}\left(\sum\limits_{t=0}^\infty \delta^t c(s(t),a(t)) | s(0)\right)$ where $\delta < 1$ is a constant discount factor, which models the fact that a higher weight is put on the current cost than the future cost. The expectation is taken over the distribution of the green power budget, the workload arrival, the environment state and the network congestion state. It is known that in MDP, this problem is equivalent to the following optimization: $\min_\pi C^\pi(s), \forall s\in \mathcal{S}$. Let $C^*(s)$ be the optimal discounted sum cost starting with state $s$. It is well-known that $\pi^*$ and $C^*(s)$ can be obtained by recursively solving the following set of Bellman equations
\begin{align}
C^*(s) = \min_{a\in \mathcal{A}}\left(c(s,a) + \delta \sum\limits_{s'\in\mathcal{S}} P(s'|s,a)C^*(s')\right), \forall s
\end{align}
In the next section, we solve this problem using the idea of dynamic programming and online learning.

\section{Post-Decision State Based Online Learning}
If all the probability distributions were known a priori, then the optimal policy  could be solved using traditional algorithms for solving Bellman equations, e.g. the value iteration and the policy iteration \cite{sutton1998reinforcement}, in an offline manner. In the considered problem, all these probability distributions are unknown a priori and hence, these algorithms are not feasible. In this section, we propose an online reinforcement learning algorithm to derive the optimal policy $\pi^*$ on-the-fly. Our solution is based on the idea of post-decision state (PDS), which exploits the partially known information about the system dynamics and allows the edge system to integrate this information into its learning process to speed up learning. Compared with conventional online reinforcement learning algorithms, e.g. Q-learning, the proposed PDS based learning algorithm significantly improves its convergence speed and run-time performance.

\subsection{Post-Decision State}
We first introduce the notion of PDS, which is the most critical idea of our proposed algorithm. In our problem, PDS is the intermediate system state after the edge system takes the computing power demand action $a(t)$ but before the green power budget $g(t)$ is realized. Specifically, the PDS in time slot $t$, denoted by $\tilde{s}(t) \triangleq (\tilde{\lambda}(t), \tilde{e}(t), \tilde{h}(t), \tilde{b}(t))$, is defined as
\begin{align}
&\tilde{\lambda}(t) = \lambda(t),~~\tilde{e}(t) = e(t), ~~\tilde{h}(t) = h(t)\\
&\tilde{b}(t) = b(t), \textrm{if}~d_{op}(\lambda(t))>b(t)\\
&\tilde{b}(t) = \max\{b(t) -d_{op}(\lambda(t)) - a(t),0\}, \textrm{otherwise}
\end{align}
As we can see, the post-decision workload state $\tilde{\lambda}(t)$, environment state $\tilde{e}(t)$ and network congestion state $\tilde{h}(t)$ remain the same because the computing power demand action $a(t)$ does not have a direct impact on these elements of the system state. The only element of the system state that may change is the battery state $b(t)$. However, it is important to notice that the post-decision battery state $\tilde{b}(t)$ is only a virtual state but not the real battery state. Given the definition of PDS, we further define the post-decision value function $V^*(\tilde{s})$ as follows:
\begin{align}
V^*(\tilde{s}) = \sum\limits_{s'\in \mathcal{S}}P(s'|\tilde{s})U^*(s')
\end{align}
where the transition $P(s'|\tilde{s})$ is now independent of the action,
\begin{align}
\tilde{P}(s|\tilde{s})
= &P_\lambda(\lambda|\tilde{\lambda})P_e(e|\tilde{e})P_h(h|\tilde{h}) \nonumber\\
&\times \sum_{g} P_g(g|\tilde{e})\textbf{1}\{b = \min\{\tilde{b} + g, B\}\}
\end{align}
For better exposition, we refer to $s$ as the ``normal'' state and $C^*(s)$ as the ``normal'' value (cost) function, in order to differentiate with their post-decision counterparts. It is obvious that $C^*(s)$ and $V^*(\tilde{s})$ are also related through:
\begin{align}\label{UV}
C^*(s) = \min_{a\in\mathcal{A}}(c(s,a) + \delta V^*(\tilde{s}))
\end{align}

The advantages of using the PDS and post-decision value function is summarized as follows.

(1) In the PDS based Bellman equations, the expectation operation is separated from the minimization operation. If we can learn and approximate the post-decision value function $V^*(\tilde{s})$, then the minimization can be solved without any prior knowledge of the system dynamics.

(2) Given $a$, the PDS decomposes the system dynamics into an a priori unknown component, i.e. $\lambda$, $e$, $h$ and $g$ whose evolution is independent of $a$, and an a priori known component, i.e. the battery state evolution is partially determined by $a$. Importantly, $\lambda$, $e$, $h$ and $g$ are also independent of the battery state $b$. This fact enables us to develop a batch update scheme on the post-decision value functions, which can significantly improve the convergence speed of the proposed PDS based reinforcement learning.

\subsection{The algorithm}
The algorithm maintains and updates a set of variables in each time slot. These variables are
\begin{itemize}
\item The one slot cost estimate $\hat{c}^t(s,a), \forall (s, a) \in \mathcal{S}\times \mathcal{A}$.
\item The post-decision value function estimate $\hat{V}^t(\tilde{s}), \forall \tilde{s} \in \tilde{\mathcal{S}}$.
\item The normal value function estimates $\hat{C}^t(s),\forall s\in \mathcal{S}$.
\end{itemize}
The superscript $t$ is used to denote the estimates at the beginning of the time slot $t$. If these estimates are accurate, i.e. $\hat{c}^t(s,a) = c(s,a)$, $\hat{V}^t(\tilde{s}) = V^*(\tilde{s})$ and $\hat{C}^t(s) = C^*(s)$, then the optimal power demand policy is readily obtained by solving \eqref{UV}. Our goal is to learn these variables over time using the realizations of the system states and costs. The algorithm works as follows: (In each time slot $t$)

\textbf{Step 1}: Determine the optimal computing power demand $a(t) = \min_a (\hat{c}^t(s(t),a) + \delta \hat{V}^t(\tilde{s}(t)))$ where for each $a$, $\tilde{s}(t)$ is the corresponding PDS. Given this power demand, the corresponding optimal offloading and autoscaling actions are determined as $\mu(t) = \mu^*(s(t), a(t))$ and $m(t) = m^*(s(t), a(t))$ based on the solution of \eqref{joint}.

After the green power budget $g(t)$ and hence the current slot cost $\tilde{c}(t)$ is realized, the battery state evolves to $b(t+1)$ according to \eqref{buffer}. The following steps update the estimates.

\textbf{Step 2}: Batch update $\hat{c}^t(s, a)$ for any action $a$ and any state $s = (\lambda,e, h, b)$ such that $e = e(t)$ using the realized green power budget $g(t)$ according to
\begin{align}
\hat{c}^{t+1}(s,a) = (1-\rho^t) \hat{c}^{t}(s,a) + \rho^t c(s, a, g(t))
\end{align}
where $\rho^t$ is the learning rate factor that satisfies $\sum_{t=0}^\infty \rho^t = \infty$ and $\sum\limits_{t=0}^\infty (\rho^t)^2 < \infty$. For all other action-state pair, $\hat{c}^{t+1}(s,a) = \hat{c}^t(s,a)$. We can do this batch update because the green power budget $g(t)$ depends only on the environment state $e(t)$ but not on other states or actions.

\textbf{Step 3}: Batch update the normal value function estimate for any state $s = (\lambda, e, h, b)$ such that $e = e(t)$ according to
\begin{align}
\hat{C}^{t+1}(s) = \min_{a\in\mathcal{A}}(\hat{c}^{t+1}(s,a) + \delta \hat{V}^t(\tilde{s}))
\end{align}
The normal value function estimates for the remaining states are unchanged.

\textbf{Step 4}: Batch update the post-decision value function estimate for any $\tilde{s}\in \tilde{\mathcal{S}}$ such that $\tilde{\lambda} = \tilde{\lambda}(t)$, $\tilde{e} =\tilde{e}(t)$ and $\tilde{h} = \tilde{h}(t)$ according to
\begin{align}
\hat{V}^{t+1}(\tilde{s}) = (1-\alpha^t) \hat{V}^{t}(\tilde{s}) + \alpha^t \hat{C}^{t+1}(s)
\end{align}
where $s = (\lambda, e, h, b)$ satisfies $\lambda = \lambda(t+1)$, $e = e(t+1)$, $h = h(t+1)$ and $b = \min\{\tilde{b} + g(t), B\}$. In this way, we update not only the currently visited PDS $\tilde{s}(t)$ but all PDS with common $\tilde{\lambda}(t)$, $\tilde{e}(t)$ and $\tilde{h}(t)$. This is because the temporal transition of $\lambda, e, h$ is independent of of the battery state $b$ and the green power budget realization follows the same distribution since the environment state $e$ is the same for these states.

\subsection{Convergence of the PDS learning algorithm}
\begin{theorem}
The PDS based online learning algorithm converges to the optimal post-decision value function $V^*(\tilde{s}), \forall \tilde{s}$ when the sequence of learning rates $\alpha^t$ satisfies $\sum_{t=0}^\infty \alpha^t = \infty$ and $\sum\limits_{t=0}^\infty (\alpha^t)^2 < \infty$.
\end{theorem}

Because $C^*(s), \forall s$ is a deterministic function of $V^*(\tilde{s}), \forall \tilde{s}$, it is straightforward that the PDS based online learning algorithm also converges to $C^*(s), \forall s$. Therefore, we prove that the edge system is able to learn the optimal power demand policy and hence the optimal offloading and autoscaling policies using the proposed algorithm.

\section{Simulation}
\begin{figure*}
  \centering
  \begin{minipage}[b]{0.31\linewidth}
  \centering
  \includegraphics[width = 1\textwidth]{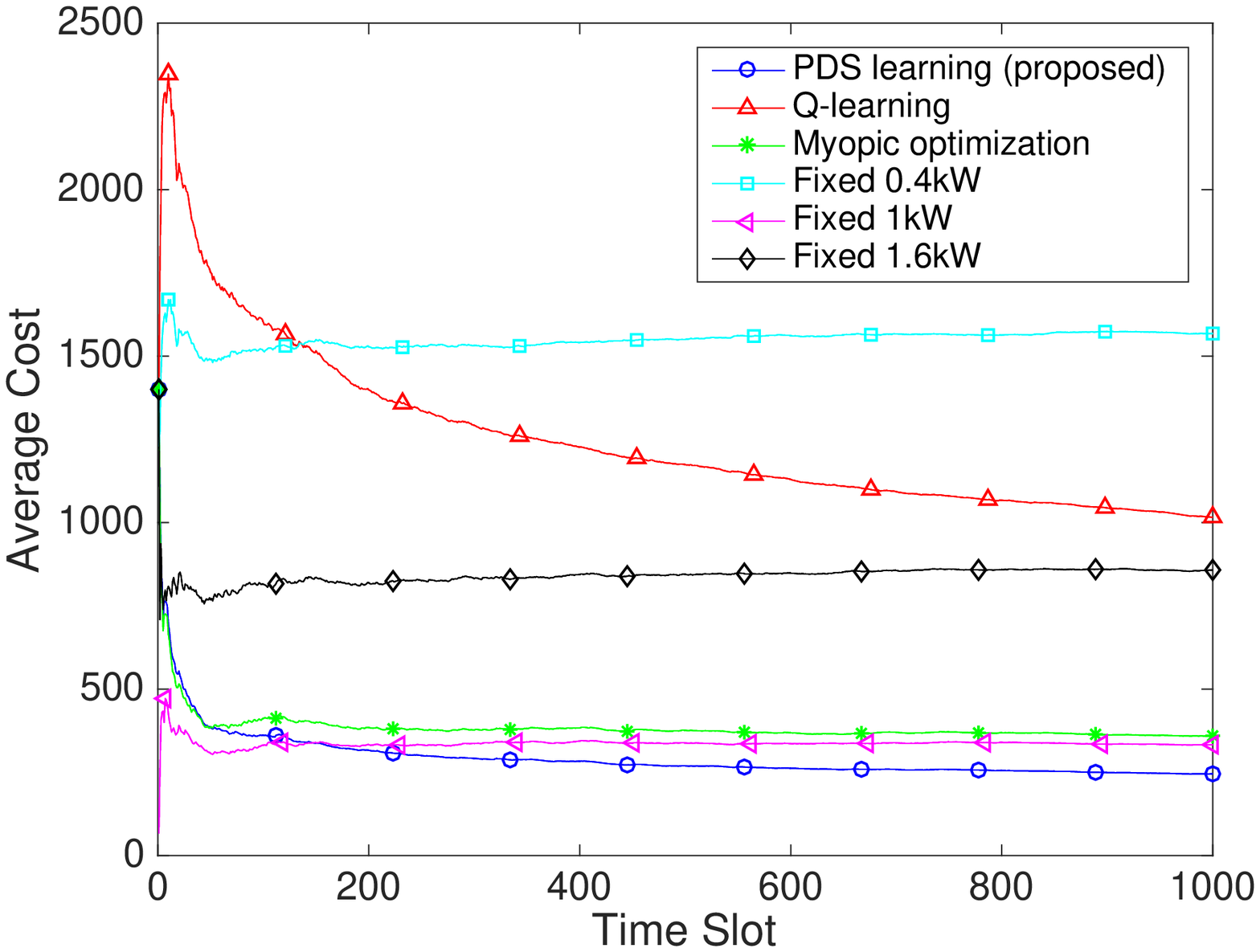}\\
  \caption{Run-time performance comparison}\label{Fig:performance}
  \end{minipage}
  \begin{minipage}[b]{0.31\linewidth}
  \centering
  \includegraphics[width = 1\textwidth]{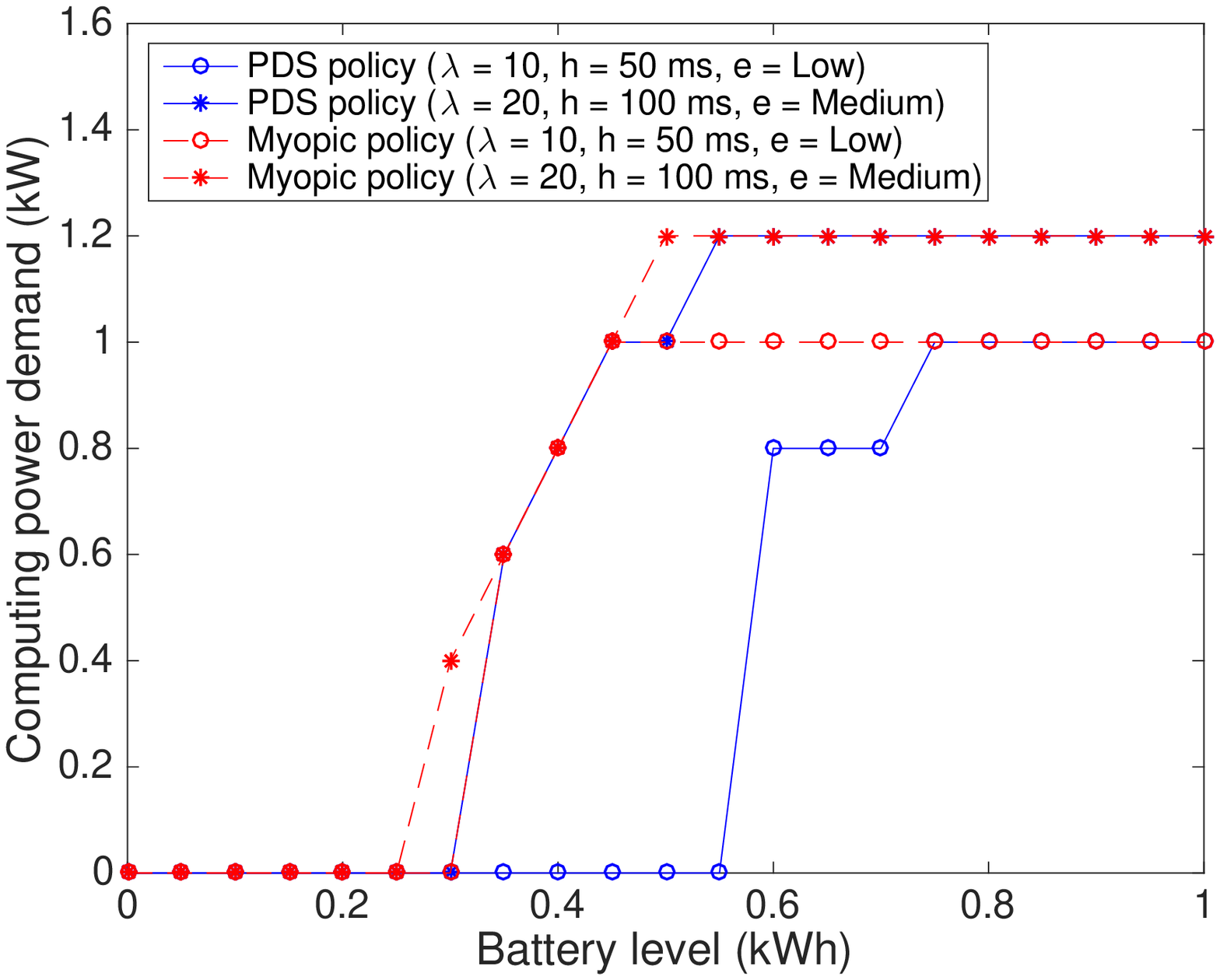}\\
  \caption{Learned computing power demand policy.}\label{Fig:policy}
  \end{minipage}
  \begin{minipage}[b]{0.31\linewidth}
  \centering
  \includegraphics[width = 1\textwidth]{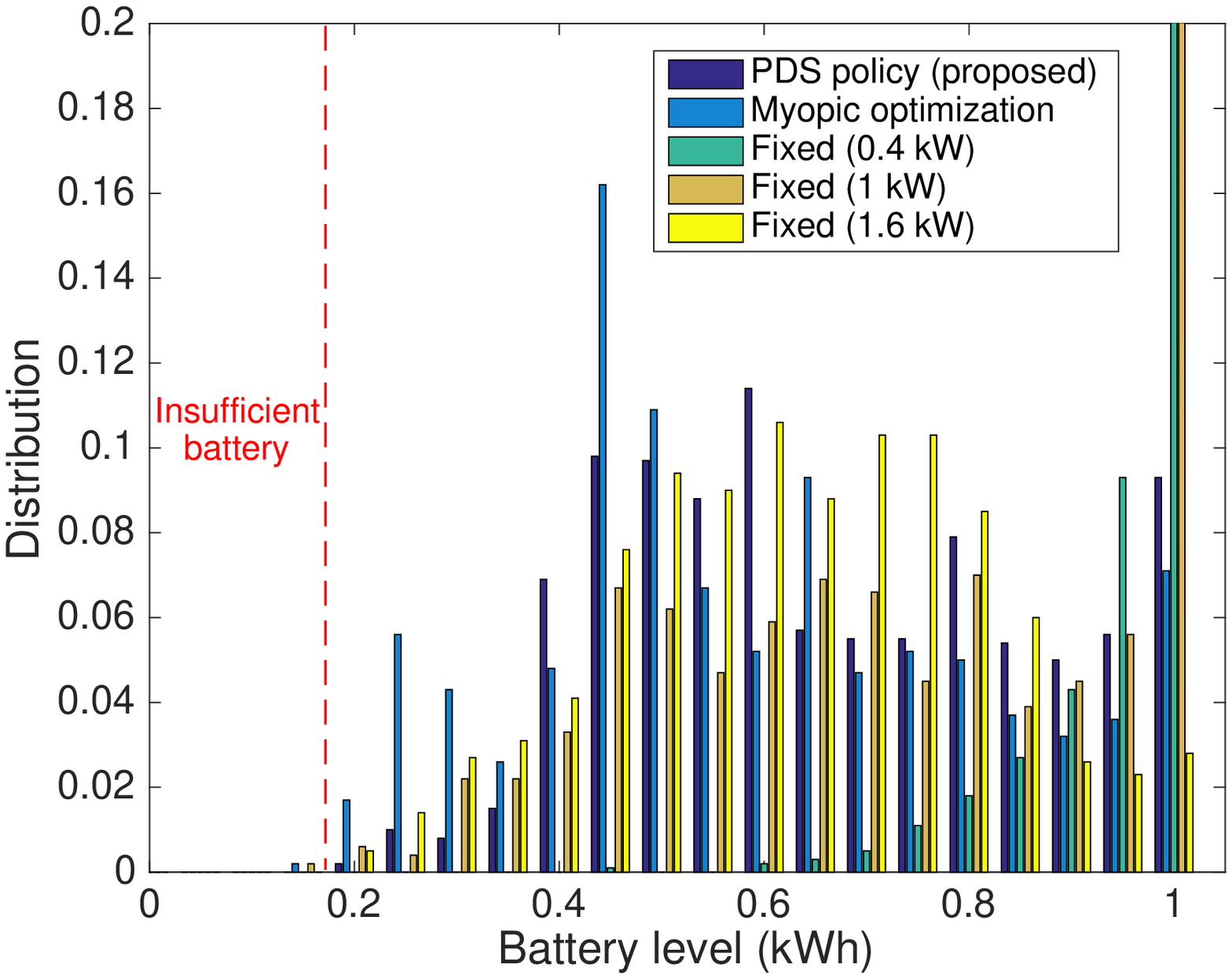}\\
  \caption{Battery state distributions.}\label{Fig:distribution}
  \end{minipage}
  \vspace{-0.1in}
\end{figure*}

We consider each time slot as 15 minutes. The workload value space is $\Lambda$=\{10 unites/sec, 20 units/sec, 30 units/sec\}. The environment state space is $E$=\{Low, Medium, High\}. For each environment state, the green power will be realized according to a normal distribution with different means. The network congestion state space is $H$=\{50ms, 200ms, 800ms\}. The battery capacity is $B$=1kWh. The base station power consumption is 800W and the power consumption of each edge server is 200W. The maximum service rate of each server is 10 units/second. We set $d_0 = 30ms$, $\omega = 0.2$ and $\phi = 10$. Three benchmark schemes are used:
{\bf Fixed power}. Fixed computing power is used whenever possible in each slot.
{\bf Myopic optimization}. This scheme ignores the temporal correlation between the system states and the decisions and minimizes the current slot cost.
{\bf Q-learning}. This is a widely-used reinforcement learning algorithm for solving MDP problems.

{\bf (1)} Figure \ref{Fig:performance} illustrates the run-time performance. Each curve is generated by averaging 30 simulation runs. Firstly, the proposed PDS-based learning algorithm incurs a significantly lower cost than all benchmark schemes. At time slot 1000, the cost reduction exceeds 25\% compared to the second-best scheme. Secondly, the fixed power schemes result in tremendously different performance, which implies that they are sensitive to system parameters. Since the system dynamics are unknown a priori and may change over time, using a fixed computing power scheme will cause significant performance loss. Thirdly, the performance of Q-learning is much worse. This is because Q-learning converges very slowly (as can be seen from the figure, there is a declining trend) due to the large state space. {\bf (2)} Figure \ref{Fig:policy} explains why the proposed algorithm outperforms the myopic solution by showing the learned policies. When the workload demand is low and the network is not congested, the policy learned by the proposed algorithm is very conservative in using local computing power. In this way, more power can be saved for future when the workload is high and the network congestion state degrades, thereby reducing the system cost in the long term. On the other hand, the myopic policy ignores this temporal correlation. It activates local servers to process workload even if the battery level is not so high. As a result, even though it achieves slight improvement in the current slot, it wastes power for potentially reducing significant cost in the future. {\bf (3)} Figure \ref{Fig:distribution} show the distribution of the battery state over 1000 slots in one representative simulation run for the various schemes. If a too small fixed power demand is used, the battery is in the high state most of the time, implying that much of the green power is wasted due to the battery capacity constraint. If a too large fixed power demand is used, the battery tends to be in the low state and hence, it is not able to support sufficiently many servers for processing a large amount of workload locally. Although a proper fixed power demand is able to strike a decent balance, it does not adapt well to the changing system dynamics. The proposed PDS-based learning algorithm achieves the highest harvesting efficiency.

\section{Conclusion}
In this paper, we studied the joint offloading and autoscaling problem in mobile edge computing systems powered by renewable energy. We found that foresightedness and adaptivity are the keys to reliable and efficient operation of renewable-powered edge computing systems. To enable fast learning in the presence of a priori unknown system parameters, a PDS-based reinforcement learning algorithm was developed to learn the optimal policy by exploiting the special structure of the considered problem.

\bibliographystyle{IEEEtran}
\bibliography{refs}

\end{document}